\documentclass[reprint,superscriptaddress,showkeys,amsmath,amssymb,aps,onecolumn]{revtex4-2}
\setcitestyle{super}    

\usepackage{graphicx}   
\usepackage{dcolumn}    
\usepackage{bm}         
\usepackage{braket}     
\usepackage{fourier}    
\usepackage{upgreek}    
\usepackage[table,xcdraw]{xcolor} 
\usepackage{orcidlink}  
\usepackage{nicefrac}   
\usepackage{ulem}

\newcommand{\sfrac}[2]{{\textstyle\frac{#1}{#2}}}

\usepackage{fancyhdr} 
\usepackage{titlesec}
\titleformat{\section}[runin]{\normalfont\bfseries\itshape}{\thesubsection}{0.5em}{}[\,\,\,--]
\titlespacing*{\section}{0pt}{*1}{5pt}

\begin{document}

\title{Electromagnetic Properties of Indium Isotopes Elucidate the Doubly Magic Character of \texorpdfstring{$\mathbf{^{100}}$}SSn}

\author{
\centering\normalsize
Karthein~J.$^{1,\diamondsuit}$\orcidlink{0000-0002-4306-9708},
Ricketts~C.M.$^2$\orcidlink{0000-0003-1084-3573},
Garcia Ruiz~R.F.$^{1,2,3,\dagger}$\orcidlink{0000-0002-2926-5569},
Billowes~J.$^2$,
Binnersley~C.L.$^{2}$\orcidlink{0000-0001-9415-1907}, 
Cocolios~T.E.$^{4}$\orcidlink{0000-0002-0456-7878},
Dobaczewski~J.$^{5,6}$\orcidlink{0000-0002-4158-3770},
Farooq-Smith~G.J.$^{4,\oplus}$\orcidlink{0000-0001-8384-7626}, 
Flanagan~K.T.$^{2}$\orcidlink{0000-0003-0847-2662},
Georgiev~G.$^7$\orcidlink{0000-0003-1467-1764},
Gins~W.$^4$\orcidlink{0000-0002-2353-7455},
de~Groote~R.P.$^{4}$\orcidlink{0000-0003-4942-1220},
Gustafsson~F.P.$^{4}$\orcidlink{0000-0003-2063-4310},
Holt~J.D.$^{8,9}$\orcidlink{0000-0003-4833-7959},
Kanellakopoulos~A.$^{4}$\orcidlink{0000-0002-6096-6304}, 
Koszor\'us~\'A.$^{4}$\orcidlink{0000-0001-7959-8786},
Leimbach~D.$^{3,10}$\orcidlink{0000-0002-4587-1067},
Lynch~K.M.$^{2,3}$\orcidlink{0000-0001-8591-2700},
Miyagi~T.$^{11,12,13,\odot}$\orcidlink{0000-0002-6529-4164}, 
Nazarewicz~W.$^{14}$\orcidlink{0000-0002-8084-7425},
Neyens~G.$^{4}$\orcidlink{0000-0001-8613-1455},
Reinhard~P.-G.$^{15}$\orcidlink{0000-0002-4505-1552},
Sahoo~B.K.$^{16}$\orcidlink{0000-0003-4397-7965},
Vernon~A.R.$^{2}$\orcidlink{0000-0001-8130-0109},
Wilkins~S.G.$^{3,\S}$\orcidlink{0000-0001-8897-7227},
Yang~X.F.$^{17}$\orcidlink{0000-0002-1633-4000},
Yordanov~D.T.$^{3,7}$\orcidlink{0000-0002-1592-7779}\\
\vspace{5pt}\small
$^1$\textit{Massachusetts Institute of Technology, Cambridge, MA, USA}\\
$^2$\textit{Department of Physics and Astronomy, The University of Manchester, Manchester, UK}\\
$^3$\textit{CERN, Geneva, Switzerland}\\
$^4$\textit{KU Leuven, Instituut voor Kern- en Stralingsfysica, Leuven, Belgium}\\
$^5$\textit{School of Physics, Engineering and Technology, University of York, UK}\\
$^6$\textit{Institute of Theoretical Physics, Faculty of Physics, University of Warsaw, Warsaw, Poland}\\
$^7$\textit{IJCLab, CNRS/IN2P3, Université Paris-Saclay, Orsay, France}\\
$^8$\textit{TRIUMF, Vancouver, BC, Canada}\\
$^9$\textit{Department of Physics, McGill University, Montr\'eal, QC, Canada}\\
$^{10}$\textit{Department of Physics, The University of Gothenburg, Gothenburg, Sweden}\\
$^{11}$\textit{Technische Universit\"at Darmstadt, Department of Physics, Darmstadt, Germany}\\
$^{12}$\textit{ExtreMe Matter Institute EMMI, GSI Helmholtzzentrum f\"ur Schwerionenforschung GmbH, Darmstadt, Germany}\\
$^{13}$\textit{Max-Planck-Institut f\"ur Kernphysik, Heidelberg, Germany}\\
$^{14}$\textit{Facility for Rare Isotope Beams \& Department of Physics and Astronomy, Michigan State University, East Lansing, MI, USA}\\
$^{15}$\textit{Institut f\"ur Theoretische Physik, Friedrich-Alexander-Universit\"at, Erlangen/N\"urnberg, Germany}\\
$^{16}$\textit{Atomic, Molecular and Optical Physics Division, Physical Research Laboratory, Navrangpura, Ahmedabad, India}\\
$^{17}$\textit{School of Physics and State Key Laboratory of Nuclear Physics and Technology, Peking University, Beijing, China}\\
$^{\oplus}$\textit{present: Department of Oncology Physics, Edinburgh Cancer Centre, Western General Hospital, Edinburgh, UK}\\
$^{\odot}$\textit{present: Center for Computational Sciences, University of Tsukuba, Ibaraki, Japan}\\
$^{\S}$\textit{present: Massachusetts Institute of Technology, Cambridge, MA, USA}\\
\vspace{5pt}\normalsize
\textit{Correspondence:}
$^\diamondsuit$\href{mailto:karthein@tamu.edu}{karthein@tamu.edu}, $^\dagger$\href{mailto:rgarciar@mit.edu}{rgarciar@mit.edu}\vspace{5pt}
}

\date{\today}


\begin{abstract} \bf
    Understanding the nuclear properties in the vicinity of $\mathbf{^{100}}$Sn – suggested to be the heaviest doubly magic nucleus with equal proton number $\mathbf{\textit{Z}}$ and neutron number $\mathbf{\textit{N}}$ – has been a long-standing challenge for experimental and theoretical nuclear physics. In particular, contradictory experimental evidence exists regarding the role of nuclear collectivity in this region of the nuclear chart. Here, we provide additional evidence for the doubly-magic character of $\mathbf{^{100}}$Sn by measuring the ground-state electromagnetic moments and nuclear charge radii of indium ($\mathbf{\textit{Z}=49}$) isotopes as $\mathbf{\textit{N}}$ approaches 50 from above using precision laser spectroscopy. Our results span almost the complete range between the two major neutron closed shells at $\mathbf{\textit{N}=50}$ and $\mathbf{\textit{N}=82}$ and reveal parabolic trends as a function of the neutron number, with a clear reduction toward these two neutron closed-shells. A detailed comparison between our experimental and numerical results from two complementary nuclear many-body frameworks, density functional theory and ab initio methods, exposes deficiencies in nuclear models and establishes a benchmark for future theoretical developments.
\end{abstract}

\maketitle



Studying magic nuclei has guided our understanding of atomic nuclei. In analogy with the filled electron shells in atomic noble gasses, these nuclear systems with filled proton or neutron shells exhibit relatively simple structures with enhanced binding and high energies of excited states~\cite{Wie13, Ste13, Gor19}. The area of the nuclear chart surrounding the $N=Z=50$ isotope $^{100}$Sn is of particular interest as this isotope has been proposed to be doubly magic with closed shells of protons and neutrons~\cite{Hin12}.

The importance of nuclear structure studies approaching $^{100}$Sn has been widely recognized~\cite{Wal94, Hin12, Dar10, Bad13, Gua13, Fae13, Cor15, Aur18, Mor18, Tog18, Zuk21, Mou21, Rep21, Nie23}. Theoretically, significant progress has been achieved in describing isotopes around $^{100}$Sn. Complementary many-body methods can now calculate ground state properties (binding energies, radii, and electromagnetic moments) of these nuclei\cite{Gor19, Mor18, Tog18, Yor20, Per21, Mou21, Ver22}. However, despite great interest, the experimental knowledge of this neutron-deficient region of the nuclear chart is still incomplete due to low production yields. In particular, contradictory experimental evidence exists on the evolution of collective properties when approaching $^{100}$Sn~\cite{Dar10, Hin12, Fae13, Ban13, Don14} (see Fig.~\ref{fig:results}a).

Here, we report results on the hyperfine structure and isotope shifts from precision laser spectroscopy experiments performed on neutron-deficient indium ($Z=49$) isotopes down to the neutron number $N=52$ and neutron-rich indium isotopes up to $N=82$. These data allowed us to extract the nuclear quadrupole moments, magnetic dipole moments, and charge radii. This makes it possible to comprehensively describe the evolution of nuclear structure towards the $N=50$ closed shell. With a single proton hole in the $Z=50$ shell, the electromagnetic properties of indium isotopes provide a compelling laboratory to study the interplay between single-particle and collective properties of nuclei between the two major neutron-closed shells $N=50$ and $N=82$. The study of indium isotopes dates back almost 100 years, where investigations of its stable isotopes provided some of the earliest indications of nuclear deformation~\cite{Sch35}.


\section{Laser Spectroscopy Measurements and Theoretical Developments}\label{sec2}

The $^{101-131}$In isotopes were produced in two separate campaigns at the Isotope Separator On-Line Device (ISOLDE) at CERN, where a 1.4$\,$GeV proton beam from CERN’s Proton Synchrotron Booster induces nuclear reactions in thick lanthanum (in the case of $^{101-115}$In) or uranium (in the case of $^{113-131}$In; see also Ref.~\cite{Ver22}) carbide targets, with the latter one employing a proton-to-neutron converter~\cite{Got14} to suppress isobaric contamination. These reactions produce a multitude of short-lived isotopes of different chemical elements. Through diffusion by heating the target to $>2000\,^{\circ}$C, these nuclides are extracted from the target and subsequently ionized via multi-step, element-selective, resonant laser ionization~\cite{Rot11} (see 304$\,$nm and 325$\,$nm laser transitions in Fig.$\,$\ref{fig:setup}). The ions are then accelerated to 39948(1)\,eV (in the case of $^{101-115}$In) or $40034(1)\,$eV\cite{Ver22} (in the case of $^{113-131}$In) and mass-selected for the isotope of interest using ISOLDE’s separator magnets.

Following cooling and bunching in ISOLDE’s radio-frequency quadrupole trap~\cite{Man09}, the ions are guided into the Collinear Resonance Ionization Spectroscopy (CRIS) setup~\cite{Ver20}. At CRIS, the arriving ion bunch is neutralized through electron exchange with sodium vapor to be left with a fast neutral beam~\cite{Ver19}. This atom beam is overlapped with lasers both temporally and spatially to induce a multi-step ionization and finally deflected onto an ion detector, resulting in nearly background-free signal detection. The Methods section in Ref.\cite{Ver22} gives more details on the setup.

The hyperfine spectrum is measured by scanning the first laser step over the 246.0\,nm and 246.8\,nm transitions (see the two right magenta laser transitions and the resulting $^{101}$In spectrum in Fig.$\,$\ref{fig:setup}). Magnetic dipole and electric quadrupole parameters, $A_{\text{hf}}$ and $B_{\text{hf}}$, as well as isotope shifts of the $I^{\pi}=9/2^+$ nuclear ground state and the $I^{\pi}=1/2^-$ nuclear-excited state (if applicable), were extracted from the measured spectra (see Methods-\ref{sec:M-A-hfs} for details).

The results are presented in Tables~\ref{tab:1} and~\ref{tab:2}. Table~\ref{tab:1} lists the spectroscopic nuclear quadrupole moments $Q_{\rm{s}}$ and the nuclear magnetic moments $\mu$ for $^{101-115}$In. The results for $^{117-131}$In are taken from Ref.~\cite{Ver22}. Table ~\ref{tab:2} presents the differential mean-square nuclear charge radii $\updelta\langle r^2\rangle$ for $^{101-131}$In. The nuclear magnetic moments $\mu$ are calculated following the description in Ref.~\cite{Ver22}.

The data collection is described in Ref.~\cite{Ver22}. Details of the data analysis for the $^{101-115}$In isotopes can be found in the Methods section~\ref{sec:M-B-fit}, and for the $^{117-131}$In isotopes in Ref.~\cite{Ver22}.

Our experimental results are compared with predictions from two complementary many-body methods, the ab initio and nuclear density functional theory (DFT). The valence-space formulation of the ab initio in-medium-similarity-renormalization-group (VS-IMSRG) calculations~\cite{Str17,Str19} (see Methods-\ref{sec:M-H-VSIMSRG} for further details) were performed using two- and three-nucleon interactions derived from chiral effective field theory~\cite{Epe09, Mac11} and are labeled “1.8/2.0(EM)”~\cite{Heb11} and “$\Delta$N$^2$LO$_{\rm{GO}}$”~\cite{Jia20}. The former is constrained by properties of two-, three- and four-nucleon systems and reproduces ground-state energies across the nuclear chart while underpredicting absolute charge radii~\cite{Mor18, Str21}. $\Delta$N$^2$LO$_{\rm{GO}}$ was recently developed to include $\Delta$-isobar degrees of freedom and is also fit to reproduce saturation properties of infinite nuclear matter, thereby improving radii predictions. Converged values for all observables discussed here were obtained with a storage scheme allowing sufficiently large inclusion of three-nucleon force matrix elements~\cite{Miy22}. Both interactions produce similar results for $Q_{\rm{s}}$ and $\updelta\langle r^2\rangle$; hence only the 1.8/2.0(EM) results are presented.

We performed DFT calculations for neutrons using the Hartree–Fock (HF; labeled "DFT: Sky(HF)") and Hartree–Fock–Bogoliubov (HFB; labeled "DFT: Sky(HFB)") approaches, i.e., single-nucleon (HF) or nucleon–hole pair excitations (HFB) as basis states with HFB introducing some pairing correlations. For protons, the HF single-hole configuration was used. Calculations were performed for the UNEDF1 Skyrme functional~\cite{Kor22} using the methodology recently developed in Refs.\cite{Sas22, Bon23d}.

We also considered other energy density functionals: a Skyrme functional SV-min~\cite{Klu09} with mixed volume and surface pairing at BCS level~\cite{Ben03}, Fy(HFB, $\Delta r$)~\cite{Rei17} a Fayans functional with additional gradient terms in surface energy and pairing~\cite{Fay00} employing full HFB, and Fy(HFB, IVP) as a Fayans functional in which proton- and neutron-pairing terms have different strengths (for details, see Methods-\ref{sec:M-I-Fayfunc} and Extended Data Tables \ref{tab:IVP1} and \ref{tab:IVP2}). All three parametrizations are fitted to the same pool of ground-state properties from Ref.~\cite{Klu09}. At the same time, the two Fayans functionals include differential radii in the calibration dataset to reproduce the isotopic trends of charge radii better. In all cases, odd-even isotopes were computed by blocking the proton $I^{\pi}=9/2^+$ orbital. The statistical uncertainties stemming from parameter calibration were estimated using linear regression~\cite{Dob14}. To simplify the presentation, we show the uncertainties only for Fy(HFB, $\Delta r$). The uncertainty bands for the other two cases are comparable. The SV-min approach and UNEDF1 Skyrme functionals, without a full treatment of pairing, give comparable results. Hence, only "DFT: Sky(HF)" and "DFT: Sky(HFB)" calculations with UNEDF1 are shown in the figures for clarity. Both DFT and ab initio calculations were carried out with bare nucleon charges and $g$-factors.


\section{Evolution of Collectivity Around $\mathbf{\textit{N\,=\,Z\,=\,50}}$}\label{sec3}

Collective properties of nuclei around $^{100}$Sn have been the focus of intense interest in nuclear physics laboratories worldwide. As the ground states of even-even tin isotopes have nuclear spin $I\,=\,0$, the evolution of their quadrupole collectivity can be studied by measuring the $B(E2)$ rate for exciting their first $2^+$ states. The experimental results for the tin isotopic chain are summarized in Fig.~\ref{fig:results}a) after conversion to spectroscopic quadrupole moments $Q_{\rm{s}}$~\cite{Zuk21} to enable direct comparison with the spectroscopic quadrupole moments of indium.

Similar to the nuclear quadrupole moments of indium\cite{Ver22}, a reduction of the $B(E2)$ values in tin towards $N=82$ indicates a reduction in collectivity~\cite{Rad04, Rad05}, thus providing evidence for the doubly magic character of $^{132}$Sn($Z=50, N=82$). However, the large uncertainties of $B(E2)$ values for neutron-deficient tin isotopes do not allow an unambiguous conclusion toward $N=50$. See also results of recent lifetime measurements in odd-$N$ systems in the region~\cite{Pas23}. Our high-precision $Q_{\rm{s}}$ results in indium establish a clear reduction in collective properties towards $N=50$, thus strongly supporting the doubly magic character of $^{100}$Sn($Z=N=50$). In fact, the indium $Q_{\rm{s}}$ at $N=50+2$ and $N=82-2$ agree well within one combined standard deviation, thereby suggesting a similar proton and a vanishing neutron core polarization at $N=50$ and $N=82$~\cite{Ver22}, the $E2$-level signature of a doubly magic shell closure\cite{CoG15, Rod20}.

While differential charge radii $\updelta\langle r^2\rangle$ generally increase on average with increasing neutron number, an additional small parabolic trend found between shell closures also provides sensitivity to nuclear deformation\cite{Gor19, Gar16, Gro20}. This parabolic trend can be emphasized by subtracting the overall linear reference. The resulting "residual" differences $\updelta\langle r^2_{\rm{res}}\rangle$, shown in Fig.~\ref{fig:results}b), were obtained by fitting the experimental data set with a parabolic function and removing the linear trend between the fit's values at $N=50$ and $N=82$ (see also Methods-\ref{sec:M-E-dr2_res} and~\ref{sec:M-F-sys}, as well as Extended Data Figures~\ref{ExtDatFig:1:dr+theory}, \ref{ExtDatFig:2:dr2-fit}, and \ref{ExtDatFig:3:shift} for details).

In both indium and tin, the reduction in residual differential radii $\updelta\langle r^2_{\rm{res}}\rangle$ towards $N=82$ indicates a decrease in collectivity, thus providing evidence for the established doubly magic character of $^{132}$Sn($Z=50, N=82$)\cite{Gor19}. However, while the extent of existing precision tin radii
(limited to $N\geq58$\cite{Gor19}) and indium radii (limited to $N\geq56$\cite{Ebe87}) prevented a conclusion toward $N=50$, our high-precision results for $\updelta\langle r^2_{\rm{res}}\rangle$ in indium down to $^{101}$In($N=52$) present a clear reduction towards $N=50$. Well-aligned with our conclusions for $Q_{\rm{s}}$, this reduction in $\updelta\langle r^2_{\rm{res}}\rangle$ also supports the doubly magic character of $^{100}$Sn($Z=N=50$).

The experimental results are compared with the theoretical predictions in Fig.~\ref{fig:results}c) for quadrupole moments, and Fig.~\ref{fig:results}d) for residual differential charge radii. Both the DFT and VS-IMSRG calculations predict parabolic trends for the quadrupole moments. However, the magnitude of $Q_{\rm{s}}$ is significantly underestimated by the VS-IMSRG calculations. This underestimation is observed in all ab initio calculations based on spherical references and results from neglected many-particle, many-hole excitations when truncated at the two- or three-body levels\cite{Str22}. We note, however, that quadrupole collectivity is well reproduced by methods based on deformed references\cite{Yao20}. The difference in the DFT predictions for quadrupole moments is small as compared to the predicted uncertainty of the calculations (see light blue error band). The (small) differences in the $Q_s$ are related to pairing strength differences. Consider, e.g., the two Fayans functionals. Fy(HFB,$\Delta r$) uses the same strength for proton and neutron pairing, thus overestimating the latter. The functional Fy(HFB, IVP) allows for a separate tuning for neutron pairing strength leading to a somewhat larger $Q_s$. This trend in $Q_{\rm{s}}$ results from pairing driving the nucleus towards a spherical shape against the increase of other deformation effects at the mid-shell~\cite{Rei84}.

All theories generally reproduce the dominating linear trend in the differential charge radii (a plot of the trend can be found, for completion, in Methods-\ref{sec:M-E-dr2_res} and in the Extended Data Fig.~\ref{fig:dr+theory}). However, the parabolic behavior (Fig.~\ref{fig:results}d) between the closed shells is reproduced by the VS-IMSRG results (although somewhat underestimated), while only the DFT-Fayans calculations with a full treatment of pairing closely approach the observed trend. DFT-Skyrme-based calculations without full pairing underestimate the curvature.

In conclusion, both the radii and quadrupole moments indicate that quadrupole deformation dominates the ground states in the indium isotopes in the neutron mid-shell region, similar to what has been indicated for Sn isotopes~\cite{Zuk21}. However, the inverse DFT behavior between $Q_{\rm{s}}$ and $\updelta\langle r^2_{\rm{res}}\rangle$ in the prediction of nuclear deformation with an increase in pairing strength highlights missing nucleus-deforming effects (e.g., quadrupole vibrations) in state-of-the-art DFT functionals. Extensive correlation studies between $Q_{\rm{s}}$ and $\updelta\langle r^2\rangle$ for different DFT parametrizations, described in more detail in Methods-\ref{sec:M-J-DFT-corr} and Extended Data Figure~\ref{ExtDatFig:5:DFT-corr}, confirm this conclusion. Meanwhile, the use of nucleon-hole pair excitations (HFB) over single-nucleon (HF) basis states in DFT, recently found to be crucial in describing nuclear magnetic moments when approaching a shell closure~\cite{Ver22}, seems to have a negligible effect on both $Q_{\rm{s}}$ and $\updelta\langle r^2_{\rm{res}}\rangle$ (green lines in Figs.~\ref{fig:results}c) and d).

The magnetic moments $\mu$ for both the $I^{\pi}=9/2^+$ ground state and the $I^{\pi}=1/2^-$ isomers continue the near-linear trend from Ref.\cite{Ebe87} when approaching $N=50$. For completion, our results are compared with literature and theoretical results in Extended Data Figure~\ref{ExtDatFig:4:mu} in Methods-\ref{sec:M-G-mu}. As recently found for neutron-rich isotopes~\cite{Miy23}, the inclusion of two-body currents in IMSRG-VS calculations is essential to describe the magnitude of \( \mu \) in neutron-deficient isotopes. While both DFT and IMSRG-VS calculations provide a relatively good description of the observed trends, a clear discrepancy can be noticed at \( N=50 \), where, in contrast to IMSRG-VS, DFT calculations predict a marked jump for $I^{\pi}=9/2^+$.

Lastly, we studied the collective potential energy surfaces (PES) from which we conclude that static triaxial deformations are irrelevant for the cases studied. Details can be found in Methods-\ref{sec:M-K-triax} and Extended Data Figure~\ref{ExtDatFig:6:triax}.


\section{Future Direct Measurements of $^\mathit{100}$Sn}\label{sec4}

Efforts to produce even more exotic neutron-deficient tin and indium isotopes are ongoing at ISOLDE. Moreover, the next-generation RIB facilities now in operation, such as FRIB in the U.S., RIKEN in Japan, and SPIRAL2 in France, will enable the study of $^{100}$Sn and its immediate neighbors.


\section*{Acknowledgements}\label{sec5}

This work was supported by ERC Consolidator Grant no. 648381 (FNPMLS); STFC grants ST/L005794/1, ST/L005786/1, ST/P004423/1, ST/M006433/1, ST/V001035/1, ST/P003885/1, and ST/V001116/1) including an Ernest Rutherford grant (ST/L002868/1); the U.S. Department of Energy, Office of Science, Office of Nuclear Physics under grants DE-SC0021176, DE-SC0023175, and DOE-DE-SC0013365; GOA 15/010 and C14/22/104 from KU Leuven, BriX Research Program No. P7/12; the FWO-Vlaanderen (Belgium); the European Unions Grant Agreement 654002 (ENSAR2); the Polish National Science Centre under Contract 2018/31/B/ST2/02220; a Leverhulme Trust Research Project Grant; the National Key R\&D Program of China (contract no. 2018YFA0404403); the National Natural Science Foundation of China (no. 11875073); the Deutsche Forschungsgemeinschaft (DFG, German Research Foundation) -- Project-ID 279384907 -- SFB 1245; the European Research Council (ERC) under the European Union’s Horizon 2020 research and innovation program (Grant Agreement No.\ 101020842); and the Natural Sciences and Engineering Research Council of Canada under grants SAPIN-2018-00027 and RGPAS-2018-522453, as well as the Arthur B. McDonald Canadian Astroparticle Physics Research Institute. JK acknowledges support from a Feodor-Lynen postdoctoral Research Fellowship of the Alexander-von-Humboldt Foundation. We acknowledge the CSC-IT Center for Science Ltd., Finland, for allocating computational resources. This project was partly undertaken on the University of York's high-performance Viking Cluster. We are grateful for computational support from the University of York High-Performance Computing service, Viking, and the Research Computing team. The VS-IMSRG calculations were performed with an allocation of computing resources on Cedar at WestGrid and The Digital Research Alliance of Canada and at the J\"ulich Supercomputing Center. BKS acknowledges use of ParamVikram-1000 HPC facility at Physical Research Laboratory (PRL), Ahmedabad, for carrying out the atomic calculations and his work at PRL supported by the Department of Space, Government of India. We would like to thank the ISOLDE technical team for their support and assistance.


\section*{Author Contributions Statement}
RFGR, JB, CLB, TEC, GJFS, KTF, WG, RPdG, AKo, KML, GN, ARV, SGW, and XFY proposed the experiment(s). CLB, TEC, GJFS, KTF, RFGR, GG, WG, RPdG, FPG, AKa, AKo, DL, KML, CMR, ARV, SGW, XFY and DY conducted the experiment(s). JK developed the Bayesian analysis code, the Gaussian process interpolation, and the Monte Carlo studies. JK, CMR, ARV, CLB, and RFGR analyzed the experimental results. JK and KML compiled the results and calculated the final values and uncertainties. JD, WN, and PGR performed theoretical (DFT) nuclear calculations and correlation studies. JDH and TM performed theoretical (VS-IMSRG) nuclear calculations. BKS performed theoretical (AR-RCC) atomic calculations. JK prepared the manuscript and figures with input from all authors, especially RFGR, TEC, JD, KTF, RPdG, JDH, AKo, KML, TM, WN, GN, PGR, BKS, and XFY.


\section*{Competing Interest Statement}
    The authors declare no competing interests.


\section*{Tables}
\newpage

\renewcommand{\tabcolsep}{1pt}
\renewcommand{\tablename}{Table} 
\renewcommand{\thetable}{1}

\begin{table*}[ht!]\centering
    \caption{Experimental results table 1. Mass number $A$, neutron number $N$, nuclear spin and parity $I^{\pi}$, m}agnetic dipole hyperfine parameter, $A_{\rm{hf}}$, nuclear magnetic moments $\mu$, electric quadrupole hyperfine parameter, $B_{\rm{hf}}$, and spectroscopic nuclear quadrupole moments $Q_{\rm{s}}$ (using the calculated electric-field gradient $\updelta^2V/\updelta z^2 = 57,600(400)\,e\cdot$fm$^2$/MHz from Ref.~\cite{Gar18}; see Methods-\ref{sec:M-B-fit} for details) for $I^{\pi}=9/2^+$ ground states and $I^{\pi}=1/2^-$ excited states (if available) of odd-$A$ indium ($Z=49$) isotopes using 246.0$\,$nm (5p $^2$P$_{\nicefrac{1}{2}}\,$→$\,$8s $^2$S$_{\nicefrac{1}{2}}$) and 246.8$\,$nm (5p $^2$P$_{\nicefrac{3}{2}}\,$→$\,$9s $^2$S$_{\nicefrac{1}{2}}$) transitions. The results are compared to literature values\cite{Ebe87} in italic font measured with the 451 nm (5p $^2$P$_{\nicefrac{3}{2}}\,$→$\,$6s $^2$S$_{\nicefrac{1}{2}}$) transition. Experimental uncertainties are in parenthesis, and theory-derived ones are in brackets.\\\it{Note: * = only obtained from one low-statistic spectrum of the 246.8-nm transition.}\label{tab:1}
\includegraphics[width=\linewidth]{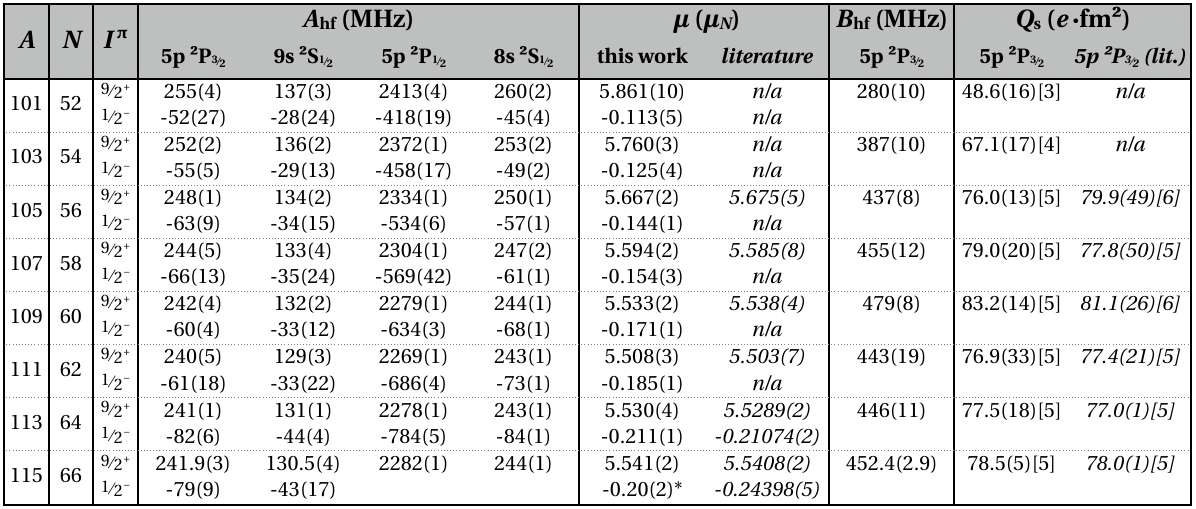}
\end{table*}

\newpage

\renewcommand{\thetable}{2}

\begin{table*}[ht!]\centering
    \caption{Experimental results table 2. Mass number $A$, neutron number $N$, nuclear spin and parity $I^{\pi}$, isotope shifts (IS) and resulting differential mean square nuclear charge radii $\updelta\langle r^2\rangle$ with respect to the centroid of the stable $^{115}$In for $I^{\pi}=9/2^+$ ground states of odd-$A$ indium ($Z=49$) isotopes for the 246.0$\,$nm (5p $^2$P$_{\nicefrac{1}{2}}\,$→$\,$8s $^2$S$_{\nicefrac{1}{2}}$) and 246.8$\,$nm (5p $^2$P$_{\nicefrac{3}{2}}\,$→$\,$9s $^2$S$_{\nicefrac{1}{2}}$) transitions (using our calculations for the field and mass shifts, and the atomic mass values from Ref.~\cite{Wan21}). Our results are compared to literature values\cite{Ebe87} in italic font measured with the 451 nm (5p $^2$P$_{\nicefrac{3}{2}}\,$→$\,$6s $^2$S$_{\nicefrac{1}{2}}$) transition. Experimental uncertainties are in parenthesis, and theory-derived ones are in brackets.}\label{tab:2}
    \includegraphics[width=.7\linewidth]{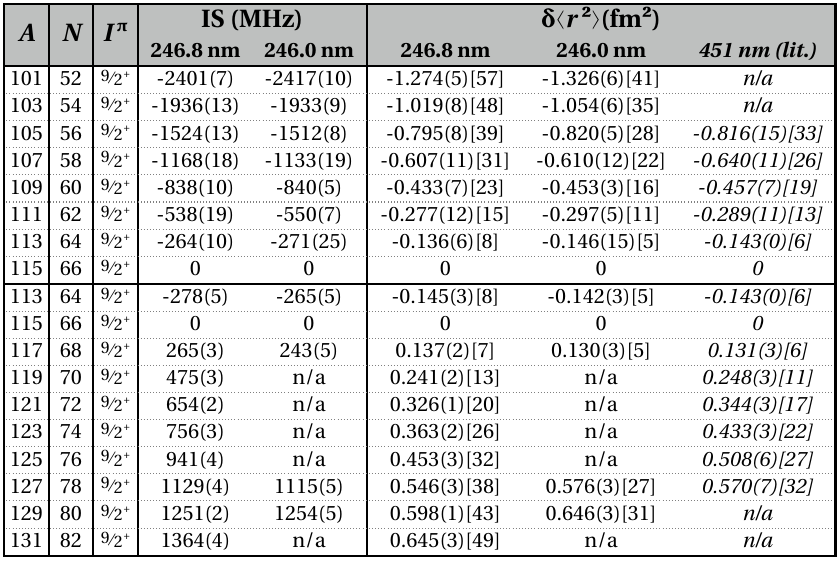}
\end{table*}


\section*{Article Figures}
\newpage

\renewcommand{\figurename}{Fig.}

\begin{figure}[h!]
    \includegraphics[width=\linewidth]{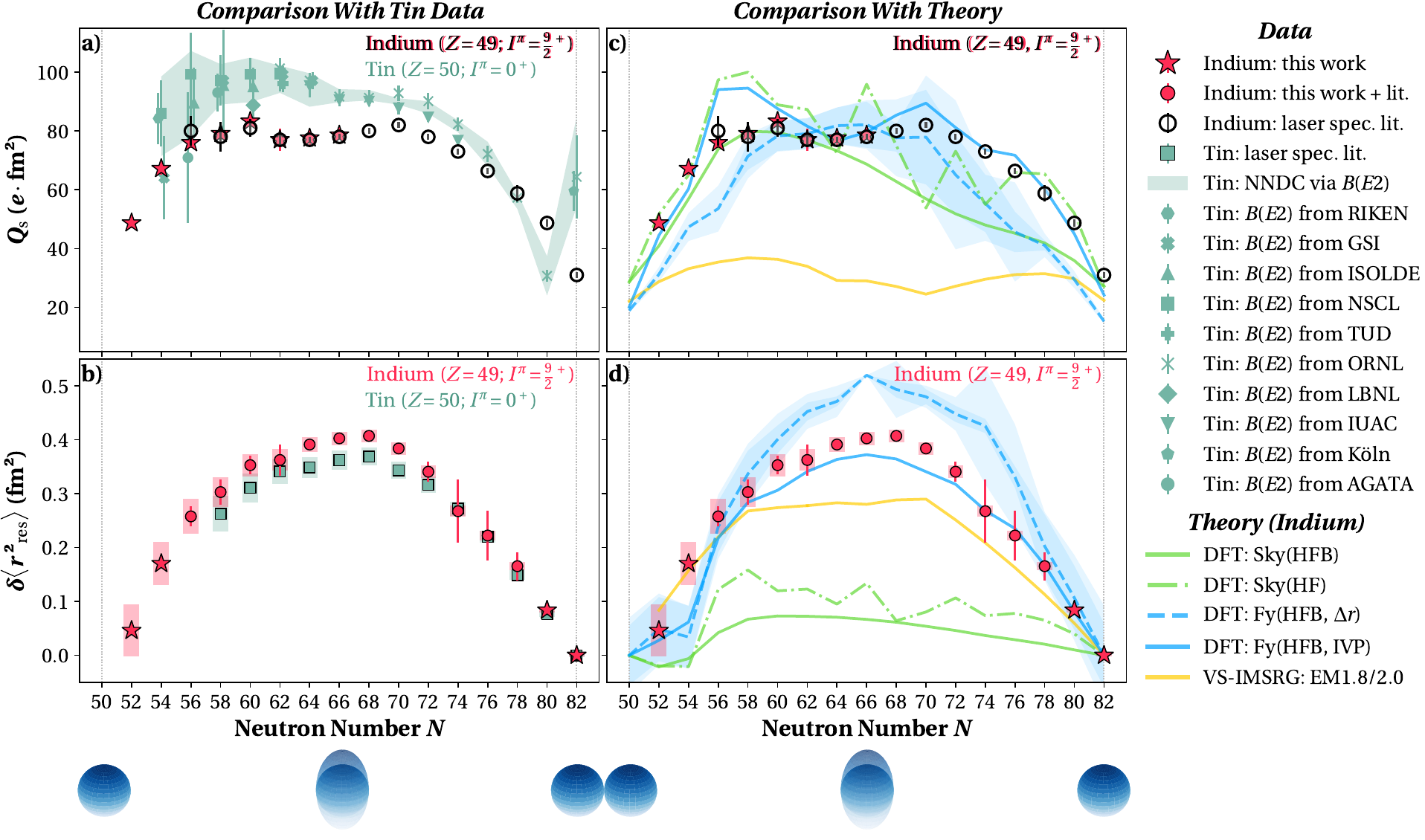}
    \caption{Evolution of the nuclear deformation between the two magic neutron numbers $N=50$ and 82 in tin and indium. Measurements in even-$N$, ground-state indium isotopes ($Z=49$; black circles = literature; red stars = this work; $I^{\pi}=9/2^+$) are compared to measurements in even-$N$, ground-state tin isotopes ($Z=50$; mint = literature; $I^{\pi}=0^+$) in the left column, and to nuclear theory (colored lines; with (darker) systematic and (lighter) statistical standard deviation bands, if applicable) in the right column. The blue spheres at the bottom represent a visual guide for the evolution of nuclear deformation between the two closed shells at $N=50$ and 82. The raw data for this figure can be found in Ref.~\cite{Kar24}.\\
    \textbf{$\,\,$--$\,\,$a/c)} Spectroscopic quadrupole moments $Q_{\rm{s}}$ with their experimental uncertainties. \textit{Indium:} Laser spectroscopy results from this work (red stars; 5p $^2$P$_{\nicefrac{3}{2}}\,$→$\,$9s $^2$S$_{\nicefrac{1}{2}}$ transition) and literature (black circles; Ref.~\cite{Ver22} for $N>74$ (same transition) and Ref.~\cite{Ebe87} for $54<N<74$ (different 5p $^2$P$_{\nicefrac{3}{2}}\,$→$\,$6s $^2$S$_{\nicefrac{1}{2}}$ transition and recalculated based on the latest atomic theory published in Ref.~\cite{Gar18})) with their experimental uncertainties; \textit{Tin:} Numerous $B(E2)$ measurements (mint borderless markers) with their standard deviations converted to spectroscopic quadrupole moments $Q_{\rm{s}}$ using Eq. 6 in~\cite{Zuk21} from RIKEN~\cite{Doo14}, GSI~\cite{Ban05, Doo08, Kum10, Gua13}, ISOLDE~\cite{Ced07, Eks08}, MSU~\cite{Vam07, Bad13}, TUD~\cite{Tog18}, ORNL~\cite{Rad04, Rad05, All11, All15}, LBNL~\cite{Kum16}, IUAC~\cite{Kum17}, Köln~\cite{Tog18}, GANIL~\cite{Sic20} and shown with their 2016 average by NNDC~\cite{NNDC16} as a mint one-sigma uncertainty band. \textit{Theory:} Selected density-functional and ab initio calculations. For details, see the text.\\
    \textbf{$\,\,$--$\,\,$b/d)} Residual differential charge radii $\updelta \langle r^2_{\rm{res}} \rangle$ with their experimental (thin; based on the weighted variance in case of indium; see Methods-\ref{sec:M-F-sys}) and theory-derived (broad) standard deviations. \textit{Indium:} Laser spectroscopy results as weighted average of the transitions from this work (red stars) and including the two recalculated transitions from Ref.~\cite{Ebe87} (red circles), if available. \textit{Tin:} Recalculated laser spectroscopy results from the 5p$^2\;^3$P$_0$→5p6s$\;^3$P$_1$ transition from Ref.~\cite{Gor19} in mint square markers. \textit{Theory:} Selected density-functional and ab initio calculations. For details, see the text and Methods-\ref{sec:M-C-atomic-theory}, \ref{sec:M-E-dr2_res}, and~\ref{sec:M-F-sys}.}
    \label{fig:results}
\end{figure}

\begin{figure}
    \includegraphics[width=\linewidth]{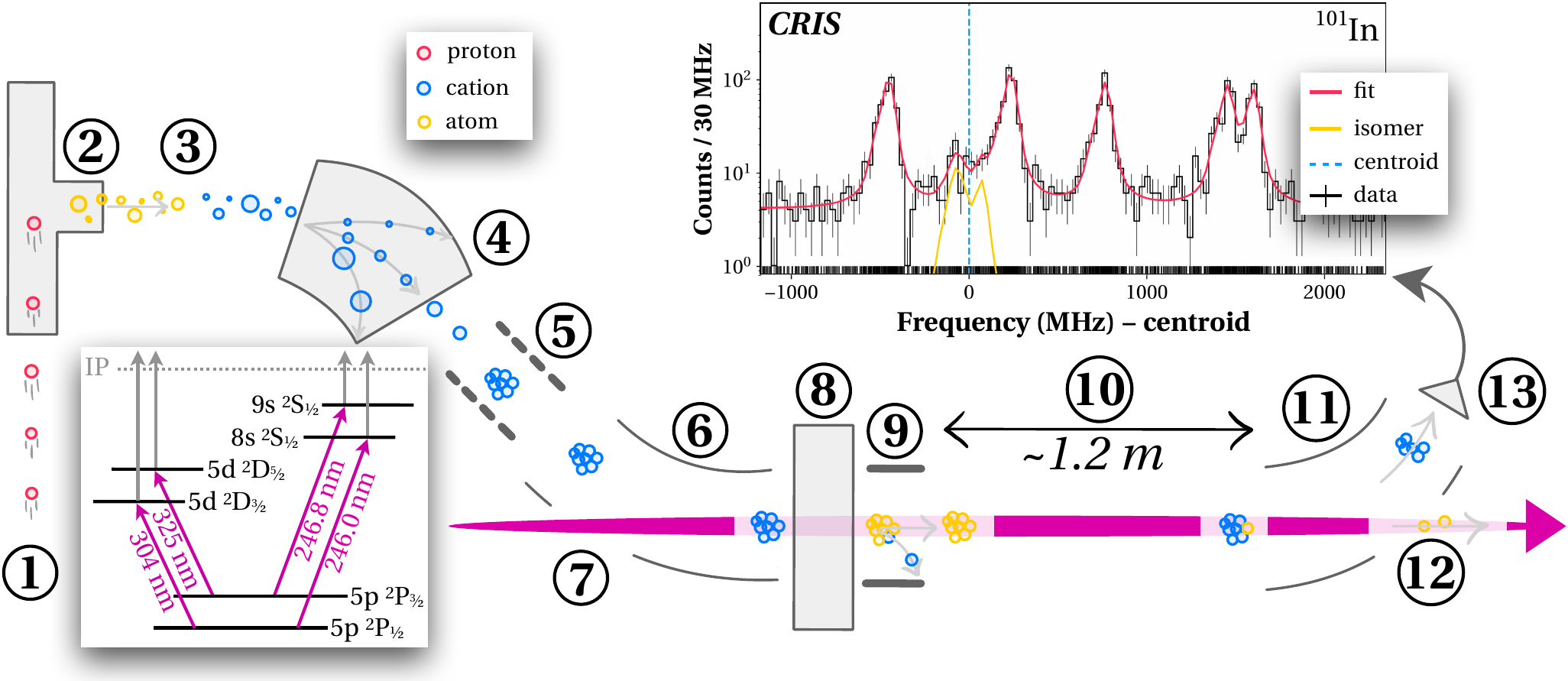}
    \caption{Schematic of the Collinear Resonance Ionization Spectroscopy (CRIS)~\cite{Ver20} experiment at CERN-ISOLDE: 1.4$\,$GeV protons (1) from CERN’s Proton Synchrotron Booster induce nuclear reactions (2) inside a lanthanum carbide target at the ISOLDE facility. Elements of interest are resonantly laser-ionized (3) beyond their ionization potential (IP) with RILIS~\cite{Rot11}, magnetic mass selected for the ion of interest (4), bunched and cooled (5) in ISCOOL~\cite{Man09}, and bent into the CRIS beamline (6). There, ions undergo neutralization in a charge-exchange cell (8) while a deflector (9) removes all remaining ions. Lasers are collinearly overlapped using laser ports (7,12) and interact with the neutral fast beam in the laser-interaction area (10; $\sim$1.2$\,$m length). Resonantly ionized ions of interest are finally bent (11) onto a particle detector (13), while background atoms remain neutral and hit the beam dump (12). The plot at the top right shows part of an example spectrum (black line with statistical standard deviation; 30$\,$MHz binning) for the $I^{\pi}=9/2^+$ ground state and $I^{\pi}=1/2^-$ isomer in $^{101}$In with a fitted theoretical line shape (red line). The inset at the bottom left shows the first resonant laser excitation transitions used for the ionization at RILIS (left two) and spectroscopy (right two). See text for details.}
    \label{fig:setup}
\end{figure}

\newpage\hbox{}\thispagestyle{empty}\newpage

\section*{Article References}

\titleformat{\section}[hang]{\large\bfseries}{\thesection}{0.5em}{}[]
\titlespacing*{\section}{0pt}{*4}{5pt}


\newpage


\appendix
\begin{center}
\textbf{\Large \noindent Methods}
\end{center}

\titleformat{\section}[runin]{\normalfont\bfseries}{\thesection}{0.7em}{|\,\,\,}[\,\,\,--]
\titlespacing*{\section}{0pt}{*1}{5pt}


\section{Hyperfine Spectrum from Raw Data}\label{sec:M-A-hfs}

The raw time-of-flight information per resonantly ionized ion was recorded on a MagneTOF\texttrademark detector by ETP Electron Multipliers Pty. Ltd. and correlated with a Doppler-shift corrected wavemeter readout of the scanning laser frequency (see the right two magenta transitions in Fig.~\ref{fig:setup}) in the rest frame of the ion at $V_0=39948(1)\,$eV (in the case of $^{101-115}$In) or $40034(1)\,$eV\cite{Ver22} (in the case of $^{113-131}$In). The time-of-flight distribution was used to remove isobaric contamination and trigger noise from the ion signal at its mass-dependent time of flight $t_i = \sqrt{\frac{2eV_0}{m_i}}$ with the elementary charge $e$. The pure hyperfine spectrum was then plotted as counts per frequency bin at the 5 MHz minimal resolution limited by the wavemeter. While the minimal bin width was typically used for fitting the spectra, studies using different bin widths up to 50 MHz were performed and used as the basis for the systematic fit uncertainty.


\section{Hyperfine Parameter and Centroid Frequencies}\label{sec:M-B-fit}

This section focuses on the analysis of $^{101-115}$In. Details on the $^{117-131}$In isotopes can be found in Ref.~\cite{Ver22}.

The hyperfine parameters, $A_{\rm{hf}}$ and $B_{\rm{hf}}$, as well as the centroid frequency, $\nu_0$, were determined through Bayesian analysis by use of binned maximum likelihood estimation. An independent analysis using the SATLAS package~\cite{Gin18} was used to confirm the validity of the code.

Due to the complex nature of the cumulative distribution function of the Voigt profile (see, e.g., Ref.~\cite{Kum20}) required for binned maximum likelihood estimation, the peak shape was estimated with a pseudo-Voigt profile using the parameters published in Ref.~\cite{Voigt}. Extensive Monte-Carlo studies were performed to verify the validity of this approximation, and deviations were found to be well within one standard deviation of the statistical uncertainty of the fit.

From all 80+ individual measurements of isotopes ranging from $^{101-115}$In, the weighted average was calculated for the hyperfine parameters per individual isotope and transition (labeled ``i''). From this data, the magnetic moments $\mu_{\rm{i}} = \frac{I_{\rm{i}}}{I_{\rm{ref}}}\cdot\frac{A_{\rm{hf}, i}}{A_{\rm{hf, ref}}}\cdot\mu_{\rm{ref}}$ and quadrupole moments $Q_{\rm{i}} = \frac{B_{\rm{hf}, i}}{e\cdot\delta^2V/\delta z^2}$ were calculated using the nuclear spin $I$ and the nuclear $E$-field gradient $\updelta^2V/\updelta z^2 = 57,600(400)\,e\cdot$fm$^2$/MHz~\cite{Gar18}. In our case, the weighted average for the stable and highly abundant $^{115}$In was used as a reference isotope (labeled ``ref'') to minimize systematic uncertainties from our measurement. In the case of the magnetic moments, the weighted average with the weighted standard deviation as its uncertainty of the four resulting values per isotope from the two studied transitions was calculated using $\mu_{\rm{ref}} = 5.5408(2) \mu_N$~\cite{Fly60}. All values are presented in Tab.~\ref{tab:1}.


\section{Atomic Calculations}\label{sec:M-C-atomic-theory}

The differential charge radii
\begin{equation*}
    \updelta\langle r^2\rangle = \frac{\nu_i-\nu_{\rm{ref}}}{F} - \frac{M}{F}\cdot\frac{m_i-m_{\rm{ref}}}{m_{i}\cdot m_{\rm{ref}}}
\end{equation*}
were calculated from the isotope shift $\Delta\nu_i = \nu_i-\nu_{\rm{ref}}$, the mass $m$ with $m_{\rm{ref}} = m(^{115}$In) or $m(^{116}$Sn) from Ref.~\cite{Hua21}, the mass shift constant $M$ and the field shift constant $F$.

For indium, we have improved calculations of isotope shift constants over our previous results~\cite{Sah20} by considering single, double, and triple excitation approximations in the analytical response coupled-cluster (AR-RCC) theory. In addition, we have included atomic orbitals belonging up to orbital angular momentum $l=6$ in contrast to our previous work where orbitals up to $l=4$ value were only considered. Corrections from the Breit and lower-order quantum electrodynamics interactions are also accounted for, improving the accuracy of the calculations. As a result, we have obtained $M=325(71)\,$GHz$\cdot$u and $F=1.577(15)\,$GHz$\cdot$fm$^{-2}$ for the 5p$^2$P$_{\nicefrac{3}{2}}$→9s$^2$S$_{\nicefrac{1}{2}}$ transition, $M=216(51)\,$GHz$\cdot$u and $F=1.626(15)\,$GHz$\cdot$fm$^{-2}$ for the 5p$^2$P$_{\nicefrac{1}{2}}$→8s$^2$S$_{\nicefrac{1}{2}}$ transition, $M=-92(74)\,$GHz$\cdot$u and $F=1.889(16)\,$GHz$\cdot$fm$^{-2}$ for the 5p$^2$P$_{\nicefrac{3}{2}}$→6s$^2$S$_{\nicefrac{1}{2}}$ transition from Ref.~\cite{Ebe87}, and $M=-158(55)\,$GHz$\cdot$u and $F=1.913(16)\,$GHz/fm$^2$ for the 5p$^2$P$_{\nicefrac{1}{2}}$→6s$^2$S$_{\nicefrac{1}{2}}$ transition from Ref.~\cite{Ebe87}.

For tin, we used $M=170(96)\,$GHz$\cdot$u and $F=2.043(113)\,$GHz$\cdot$fm$^{-2}$ from a recent analysis in Ref.~\cite{Par21} for the 5p$^2\;^3$P$_0$→5p6s$\;^3$P$_1$ transition from Ref.~\cite{Gor19}.


\section{Gaussian Process Interpolation}

We employed a Gaussian process interpolation algorithm for the isotope shifts to establish a continuous time evolution of the reference centroid frequencies $\nu_{\rm{ref}}$ also at the point in times of the measurement of an ion of interest $\nu_i$. This step is necessary to avoid systematic shifts in the isotope shifts since the individual frequency measurements using the wavemeter are dependent on environmental conditions such as room temperature. Moving to Gaussian processes for this interpolation has recently led to success across a broad spectrum of physics~\cite{GaussianProcess}. Most importantly, this technique removes all priors with respect to the choice of the temporal distribution of the frequencies and provides a statistical uncertainty for the reference isotope at the median time of the measurement of the ion of interest, which can be carried forward to the calculation of the differential charge radii $\updelta\langle r^2\rangle$. The final isotope shift values are in Tab.~\ref{tab:2}.

\setlength{\textfloatsep}{12pt}

\section{Residual Differential Charge Radii}\label{sec:M-E-dr2_res}
We isolated the parabolic trend of the measured differential charge radii with $N$ by subtracting the linear component of the $\updelta\langle r^2\rangle$ trend to study the subtle effects of the nuclear deformation and pairing. These “residual” changes $\updelta\langle r^2_{\rm{res}}\rangle$ were obtained by fitting the experimental data for the isotopic chains of indium and tin, each with a parabolic function $fit(N)=\updelta\langle r^2\rangle_{N=82}+b\cdot(N-82)+c\cdot(N-50)\cdot(N-82)$ (see Extended Data Figure$\,$\ref{ExtDatFig:2:dr2-fit}). This function was fixed to the experimental differential charge radii at $N=82$, $\updelta\langle r^2\rangle_{N=82}$. The linear trends between the fit’s values at $N=50$ and $N=82$ were then subtracted from all differential charge radii per isotope.
We also note that semi-empirical models, such as the Zamick-Talmi model\cite{Zam71}, have been proposed to explain the quadratic increase in nuclear radii between neutron closed shells.

Due to slight systematic shifts observed between the calculated charge radii of different transitions as described in Methods-\ref{sec:M-F-sys} and Extended Data Figure~\ref{ExtDatFig:3:shift}, the weighted average of all existing transitions in indium (the two from this work (red stars in Extended Data Figure~\ref{ExtDatFig:1:dr+theory}) and in addition, the two recalculated transitions from Ref.~\cite{Ebe87}  (red circles in Extended Data Figure~\ref{ExtDatFig:1:dr+theory}) using the latest atomic theory (see Methods-\ref{sec:M-C-atomic-theory}), if available, and their weighted variance as experimental uncertainty were used for these shell-to-shell fits and thus for all presented $\updelta\langle r^2\rangle$ and $\updelta\langle r^2_{\rm{res}}\rangle$ values in the figures of this article.

The theoretical uncertainty of this fit contains two contributions. First, a Monte-Carlo simulation was performed using 10000 random pairs of $F$ and $M$ from a uniform distribution within the ranges of the values’ theoretical uncertainties. For each random pair of $F$ and $M$ values, a set of $\updelta\langle r^2\rangle$ was calculated from the measured isotope shifts per transition and fitted with the same parabolic function. Next, the linear trend between $N=50$ and $N=82$ of each fit was used to calculate a set of $\updelta\langle r^2_{\rm{res}}\rangle$ per transition.

Finally, the standard deviation over all 10000 sets of $\updelta\langle r^2_{\rm{res}}\rangle$ for each isotope was calculated. This quantity allowed us to determine the uncorrelated contribution of the highly correlated theoretical uncertainties on $F$ and $M$. This correlation is reflected in a primary rotation of the trend in $\langle\delta r^2\rangle$ when $F$ and $M$ values change for the same experimental isotope shifts of a single transition. Hence, when subtracting the linear components for two sets of $F$ and $M$, their slope would change, but the residual parabola would remain except for a minuscule change in curvature. In other words, the large theoretical uncertainties from the $F$ and $M$ values are mostly removed when calculating $\updelta\langle r^2_{\rm{res}}\rangle$ due to their highly correlated nature.

This effect is captured in our Monte Carlo studies, in which we primarily find a (small) standard deviation in $\updelta\langle r^2_{\rm{res}}\rangle$ at the central data points (= change in curvature) and a vanishing standard deviation toward the closed shells (= fixed in the process to $\updelta\langle r^2_{\rm{res}}\rangle$ = 0).

The second uncertainty contribution captures the fact that there is no isotope shift measurement for $N=50$ yet; thus, this value is interpolated with this parabolic fit. This uncertainty is numerically propagated from the Jacobi matrix of the parabolic fit of the differential charge radii using their experimental uncertainties. This fit uncertainty is finally inflated by the square root of the reduced chi-square $\sqrt{\chi^2_{\rm{red}}}$ of this fit to account for the imperfect representation of our data by this simple parabolic model. This uncertainty is carried forward to calculating $\updelta\langle r^2_{\rm{res}}\rangle$. Finally, both uncertainty contributions were added in quadrature.

For theoretical calculation, these so-called ``residual'' differential charge radii $\updelta\langle r^2_{\rm{res}}\rangle$ were derived from the individual linear trends spanning between $N=50$ and $N=82$.


\section{Differential Charge Radii Comparison with Literature}\label{sec:M-F-sys}

Globally speaking, atomic theory has made tremendous progress in recent years to allow for comparisons of differential charge radii of different transitions~\cite{Sah20}. In the case of indium $I^{\pi}=9/2^+$ ground states, results for four different transitions are now available and agree well within their combined uncertainties (see Extended Data Figure~\ref{ExtDatFig:3:shift}a).

With a closer look at Extended Data Figure~\ref{ExtDatFig:3:shift}b), a slight global systematic shift can be found between different transitions. This systematic shift, in principle, is covered by sizeable theoretical uncertainties but highlights the need for further theoretical advances, particularly for the most exotic isotopes, as the deviations grow.

Moreover, a $>1\sigma$ deviation at $N=74$ and $N=76$ was found between the experimental values from the two transitions from Ref.~\cite{Ebe87} and our results. While a measurement has to be performed to clarify this deviation between measurements, we decided to present instead the weighted average between all four transitions (the two presented here and the two, where available, from Ref.~\cite{Ebe87}) to compare our results in indium with ones in tin and theoretical calculations in indium. As experimental uncertainty, we display the square root of the weighted variance to account for the deviation between the measurements, in particular at $N=74$ and $N=76$. This procedure was used in Extended Data Figures~\ref{ExtDatFig:1:dr+theory} and~\ref{ExtDatFig:2:dr2-fit} for the differential charge radii $\updelta\langle r^2\rangle$, as well as in Fig.~\ref{fig:results} and Extended Data Figure~\ref{ExtDatFig:2:dr2-fit} for the residual differential charge radii $\updelta\langle r^2_{\rm{res}}\rangle$ explained in the previous Methods-\ref{sec:M-E-dr2_res}.

Please note that while the linear behavior subtracted for the residual differential charge radii was derived from the parabolic fit of the weighted average of the differential charge radii in indium, the Monte-Carlo-derived theoretical uncertainty was calculated from the mean standard deviation of the individual transitions since the simulation was distributed over the transition-dependent $F$ and $M$ values, as described in more details in the previous Methods-\ref{sec:M-E-dr2_res}.


\section{Magnetic Moments} \label{sec:M-G-mu}
For completion, we provide a plot of the magnetic moments presented in Tab.~\ref{tab:1}. Our results for the $I^{\pi}=9/2^+$ ground states and $I^{\pi}=1/2^-$ isomers shown in Extended Data Figure~\ref{ExtDatFig:4:mu} continue the near-linear increase from Ref.\cite{Ebe87} when approaching $N=50$ and agree with our DFT results of the Hartree-Fock basis (see Ref.\cite{Ver22} for details). Our VS-IMSRG calculations \cite{Miy23b} show that the inclusion of two-body operators (labeled "1B+2B") improves the agreement with the experiment compared to results that only use one-body operators (labeled "1B").


\section{Valence-Space In-Medium Similarity Renormalization Group} \label{sec:M-H-VSIMSRG}

The valence-space in-medium similarity renormalization group (VS-IMSRG)~\cite{Her16, Str19} is an ab initio many-body method starting from two-plus-three-nucleon interactions expressed within the harmonic oscillator basis. In particular, we use two state-of-the-art interactions derived within the context of chiral effective field theory, labeled "1.8/2.0 (EM)"~\cite{Heb11,Sim17} and "$\Delta$N$^{2}$LO$_{\rm GO}$"~\cite{Jia20}.

The calculations were performed in the 15 major HO shells with frequency $\hbar\omega = 16$ MeV. For three-nucleon matrix elements, an additional truncation $E_{\rm 3max}$ defined as the sum of the three-body HO quanta needs to be introduced, and sufficiently large $E_{\rm 3max}=24$ was used with recently introduced storage scheme~\cite{Miy22}. After transforming to the spherical Hartree-Fock basis, we solve the VS-IMSRG flow equation at the two-body level of approximation, IMSRG(2), and obtain an approximate unitary transformation~\cite{Mor15} such that a targeted valence space is decoupled from the full many-body space.  We furthermore use ensemble normal ordering to capture the effects of three-nucleon forces between valence particles~\cite{Str17}. The same unitary transformation is applied for the radius, magnetic dipole, and electric quadrupole operators to compute the radius and electromagnetic moments consistently\cite{Par17}. We note that while magnetic moments are generally well reproduced, albeit with effects of two-body currents still neglected. However, within a spherical reference, many-body correlations at the two- or even three-body level are typically insufficient to fully capture quadrupole collectivity\cite{Hen18,Str22}.

In this work, we use the multi-shell VS-IMSRG\cite{Miy20} to decouple the $\{1p_{1/2}$, $1p_{3/2}$, $0f_{5/2}$, $0g_{9/2}\}$ and neutron $\{2s_{1/2}$, $1d_{3/2}$, $1d_{5/2}$, $0g_{7/2}$, $0h_{11/2}\}$ valence space above the $^{78}$Ni core, the same space as used in the previous work for neutron-rich indium isotopes~\cite{Ver22}. The two- and three-nucleon HO matrix elements were computed with the \texttt{NuHamil} code~\cite{Miy23}, the VS-IMSRG step was performed with the \texttt{imsrg++} code~\cite{imsrg++}, and the subsequent valence-space problem was exactly solved with the \texttt{KSHELL} code~\cite{KSHELL}.


\section{The Fayans Functional Fy(HFB, IVP)} \label{sec:M-I-Fayfunc}

The most widely used non-relativistic DFT calculations use the Skyrme functional~\cite{Ben03a}. An interesting extension is the Fayans functional, which was proposed in Refs.~\cite{Fay94, Kro95, Fay98, Tol15} and received renewed attention~\cite{Rei17, Mil19}. It differs from the Skyrme functional in three aspects: density dependence in terms of rational approximations rather than power-law; an additional gradient term in the surface energy; and a gradient term in the pairing functional. The latter two terms provide increased flexibility for describing isotopic differences in charge radii. The Fayans parametrization Fy(HFB, $\Delta r$) was presented in Ref.~\cite{Mil19}. The parametrization Fy(HFB, IVP) presented here goes one step further by making proton and neutron pairing terms different. Below, we briefly provide the functional and its parameters.

The kinetic energy and Coulomb Hartree terms of Fayans EDF are exactly the same as in the Skyrme model. The Fayans interaction functional is usually written in terms of dimensionless densities
\begin{equation}
    x_t=\frac{\rho_t}{\rho_{\rm sat}}\;,
\end{equation}
where $t\in\{+,-\}$ and $\rho_{\rm sat}$ is a scaling parameter of the Fayans functional, which characterizes the saturation density of the symmetric nuclear matter with Fermi energy $\varepsilon_F=(9\pi/8)^{2/3}\hbar^2/2mr_s^2$ and the Wigner-Seitz radius $r_s= (3/4\pi\rho_{\rm sat})^{1/3}$. Note that we have to distinguish between the parameter $\rho_{\rm sat}$, which is a fixed input to the model, and the equilibrium density $\rho_{\rm eq}$, which is the result of the calibration process and characterizes the Fayans functional (see below).

The Fayans functional, in a similar fashion to the Skyrme functional, is expressed in terms of the local densities $\rho_\pm=\rho_p\pm\rho_n$, kinetic densities $\tau_\pm$, spin-orbit densities $J_\pm$, and pairing densities $\breve\rho_{p/n}$. We use it in the form of FaNDF$^0$ of Ref.\cite{Fay98} starting from the total energy as
\begin{subequations}
\label{eq:Fyfunc}
\begin{eqnarray}
  {E}
  & = &
  {E}_\mathrm{kin}
      +\int d^3r\mathcal{E}_\mathrm{Fy}(\rho,\tau,\vec{j},\vec{J})
\nonumber\\
  &&
   +{E}_\mathrm{C}[\rho_C]
   -{E}_\mathrm{cm}
\label{eq:Etot} \\
   {E}_\mathrm{kin}
   &= &
   \int d^3r\left[\frac{\hbar^2}{2m_p} \tau_p+\frac{\hbar^2}{2m_n} \tau_n\right]
\label{eq:Ekin} \\
\mathcal{E}_\mathrm{Fy}
   &=&
   \mathcal{E}_\mathrm{Fy}^\mathrm{(vol)}(\rho)
   +\mathcal{E}_\mathrm{Fy}^\mathrm{(surf)}(\rho)
   +\mathcal{E}_\mathrm{Fy}^\mathrm{(ls)}(\rho,\vec{J})
\nonumber\\
   &&
   +\mathcal{E}_\mathrm{Fy}^\mathrm{(pair)}(\rho,\breve\rho)
\label{eq:EFay}\\
  \mathcal{E}_\mathrm{Fy}^\mathrm{(vol)}
  &=&
  \sfrac{2}{3}\epsilon_F^0\rho_0
  \Big[
  a_+^\mathrm{v}
  \frac{1-h_{1+}^\mathrm{v}x_+^{\sigma}}
    {1+h_{2+}^\mathrm{v}x_+^{\sigma}}x_+^2
\nonumber\\
   &&\qquad
  +
  a_-^\mathrm{v}
  \frac{1-h_{1-}^\mathrm{v}x_+}
       {1+h_{2-}^\mathrm{v}x_+}x_-^2
  \Big]
\label{eq:EFay-dens}
\end{eqnarray}

\begin{eqnarray}    \mathcal{E}_\mathrm{Fy}^\mathrm{(surf)}
  &=&
  \sfrac{2}{3}\epsilon_F^0\rho_0\Big[
  \frac{a_+^\mathrm{s}r_0^2(\nabla x_+)^2}{1+h_{+}^\mathrm{s}x_+^{\sigma}+h_{\nabla}^\mathrm{s}r_0^2(\nabla x_+)^2}
  \Big]
\label{eq:EFay-grad}
\\
  \mathcal{E}_\mathrm{Fy}^\mathrm{(ls)}
  &=&
  C_0^{(\nabla J)}\rho\vec{\nabla}\cdot\vec{J}
  +
  C_1^{(\nabla J)}\rho_-^{\mbox{}}\vec{\nabla}\cdot\vec{J}_-^{\mbox{}}
  \;,
\label{eq:ls}
\end{eqnarray}
\begin{eqnarray}
  \mathcal{E}_\mathrm{Fy}^\mathrm{(pair)}
  &=&
  \frac{2\epsilon_F^0}{3\rho_0}
  \Big[(f_\mathrm{ex}^\mathrm{(\xi)}\!+\!\delta f^\mathrm{(\xi)})\breve\rho^2_p
  +
  (f_\mathrm{ex}^\mathrm{(\xi)}\!-\!\delta f^\mathrm{(\xi)})\breve\rho^2_n
\nonumber\\
  &&\;\;
   +
    \left({h}_{1}^{(\xi)}x_+^{\gamma}
       +{h}_{\nabla}^{(\xi)}r_0^2(\nabla x_+)^2\right)
   (\breve\rho^2_p\!+\!\breve\rho^2_n)
  \Big]
  \;,
\label{eq:ep2}
\\
  \epsilon_F^0
  &=&
  \left(\frac{9\pi}{8}\right)^{2/3}\,\frac{\hbar^2}{2mr_0^2}
  \;,\;
  r_0
  =
  \left(\frac{3\pi}{8}\rho_0\right)^{1/3}.
\end{eqnarray}
\end{subequations}

Following the original FaNDF$^0$ definitions~\cite{Fay98}, we use $\hbar^2/2m_p=20.749811$\,MeV\,fm$^2$, $\hbar^2/2m_n=20.721249$\,MeV\,fm$^2$, $e^2=1.43996448$\,MeV\,fm, $\rho_{\rm sat}=0.16$\,fm$^{-3}$, $\gamma=2/3$, and $\sigma=1/3$. The Coulomb energy ${E}_\mathrm{C}$ consists out of the direct (Hartree) term with the charge density $\rho_C$ and exchange treated in Slater approximation employing $\rho_p$. The center-of-mass correction is subtracted a posteriori with $E_\mathrm{cm}=\langle\hat{P}_\mathrm{cm}^2\rangle/(2m)$ using an average nucleon mass $m$. The model parameters for Fy(HFB, IVP) are given in Extended Data Table~\ref{tab:IVP1}, and the nuclear matter properties of this functional are shown in Extended Data Table~\ref{tab:IVP2}. The latter quantifies the physical volume properties of the functional and is useful for comparison with other nuclear models.

Pairing deserves special attention because the pairing functional (\ref{eq:ep2}) alone does not fully determine the results. The size of the pairing space also plays a role. We employ a soft single-particle (s.p.) energy cutoff
\begin{equation}
  w_\alpha^\mathrm{(cut)}
  =
  \frac{1}{1+\exp\left(\frac{\varepsilon_\alpha-(\epsilon_F+\epsilon_\mathrm{cut})}{\epsilon_\mathrm{cut}/10}\right)}
\end{equation}
including all s.p. states $\alpha$ up to about $\epsilon_\mathrm{cut}=15$ MeV above the Fermi energy $\epsilon_F$. The s.p. pairing gap is modified as $\Delta_\alpha\longrightarrow\Delta_\alpha w_\alpha^\mathrm{(cut)}$ to turn off the contributions from all s.p. states above the pairing band. A further problem is that nuclear pairing is generally weak and often approaches the critical point of the pairing phase transition. This holds particularly for the parametrization Fy(HFB, IVP). The proton number $Z=49$ of In isotopes is next to the magic number at $Z=50$, aggravating the phase transition problem. Thus we use pairing stabilization as described in Ref.~\cite{Erl08} with a stabilizing lower bound of pairing energy of 0.3 MeV. The parameters in Extended Data Tables~\ref{tab:IVP1} and~\ref{tab:IVP2} were fitted in connection with that particular pairing treatment.


\section{Correlation Studies for Different DFT Parametrizations}
\label{sec:M-J-DFT-corr}

In order to improve the description of nuclear charge radii, the DFT parametrization Fy(HFB, IVP) was developed in this work (see Methods-\ref{sec:M-I-Fayfunc} and Extended Data Tables~\ref{tab:IVP1} and~\ref{tab:IVP2} for details). To study the interplay between charge radii and deformations, we inspected the correlation between the nuclear quadrupole moments and charge radii for three DFT parametrizations. The corresponding error ellipses are shown in Extended Data Figure~\ref{ExtDatFig:5:DFT-corr} together with experimental values. All three predictions agree within their error ellipses with the experiment and with each other. While the Fy(HFB, IVP) model with an isovector pairing term shows a strong correlation between $Q_{\rm{s}}$ and $\updelta\langle r^2\rangle$, no such correlation was found for the other models.


\section{Study on Triaxiality}\label{sec:M-K-triax}

Neutron-poor isotopes in the Cd-In-Sn region are known to have soft potential-energy surfaces (PES) in the landscape of quadrupole deformations. This raises the question of the potential triaxiality of these nuclei, which will be explored in this section.

Extended Data Figure~\ref{ExtDatFig:6:triax} shows the PES along axially symmetric quadrupole deformation for selected In and Cd isotopes. The even-even Cd isotopes with $N<76$ have extremely soft PES without well-developed deformed minima. Such a situation is prone to triaxiality. Thus we further checked for triaxial minima in the even-even Cd isotopes using the fully 3D DFT code of \cite{Maruhn2014} and found none. The right panel of Extended Data Figure~\ref{ExtDatFig:6:triax} shows the PES for the $I^{\pi}=9/2^+$ configuration in odd-even In isotopes. This PES exhibits a quite different behavior, with the odd proton exerting a strong quadrupole polarization which gives rise to a well-defined axial-prolate minimum for all isotopes. Such a more rigid PES gives no leeway to escape to a triaxial configuration. Thus, we conclude that static triaxial deformations are irrelevant for the cases studied.


\section*{Data Availability}

The full dataset of the $^{101-115}$In isotopes is available in Ref.~\cite{Kar23b}, and of the $^{113-131}$In isotopes in Ref.~\cite{Ver22b}. An example spectrum for the previously unmeasured $^{101}$In isotope, most relevant to this work, is included in Fig.~\ref{fig:setup} of this article.

\section*{Code Availability}
Both data repositories, Ref.~\cite{Kar23b} for $^{101-115}$In and Ref.~\cite{Ver22b} for $^{113-131}$In, contain Python scripts to read the raw data, which can be analyzed using code described in detail in Ref.~\cite{Gin18, Gin17}. Data analysis used NumPy \cite{numpy}, SciPy \cite{scipy}, Pandas \cite{pandas1, pandas2}, and iMinuit \cite{iminuit1, iminuit2}. Figures were produced using Matplotlib \cite{matplotlib}.

\section*{Extended Data Tables}

\newpage

\renewcommand{\tablename}{Ext. Data Table} 
\renewcommand{\thetable}{1}
\begin{table}
    \caption{Parameters of the Fayans functional Fy(HFB, IVP) as given formally in Eqs. (\ref{eq:Fyfunc}). The parameters are dimensionless by virtue of their scaling factors, which are quantitatively fixed by setting $\rho_0=0.08$ fm$^{-3}$.}
    \includegraphics[width=0.3\linewidth]{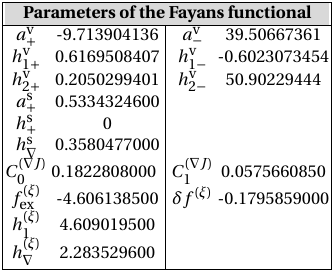}
    \label{tab:IVP1}
\end{table}

\renewcommand{\thetable}{2}
\begin{table}
    \caption{Properties of symmetric nuclear matter at its equilibrium point for the Fayans parametrization Fy(HFB, IVP). For a definition of the nuclear matter parameters see, e.g., \cite{Klu09}.}
    \includegraphics[width=0.3\linewidth]{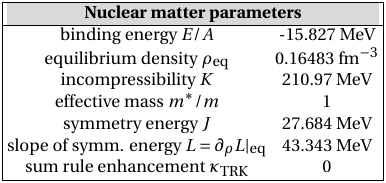}
    \label{tab:IVP2}
\end{table}

\section*{Extended Data Figures}

\newpage

\renewcommand{\figurename}{Ext. Data Fig.}
\renewcommand{\thefigure}{1}

\begin{figure}
    \includegraphics[width=0.5\linewidth]{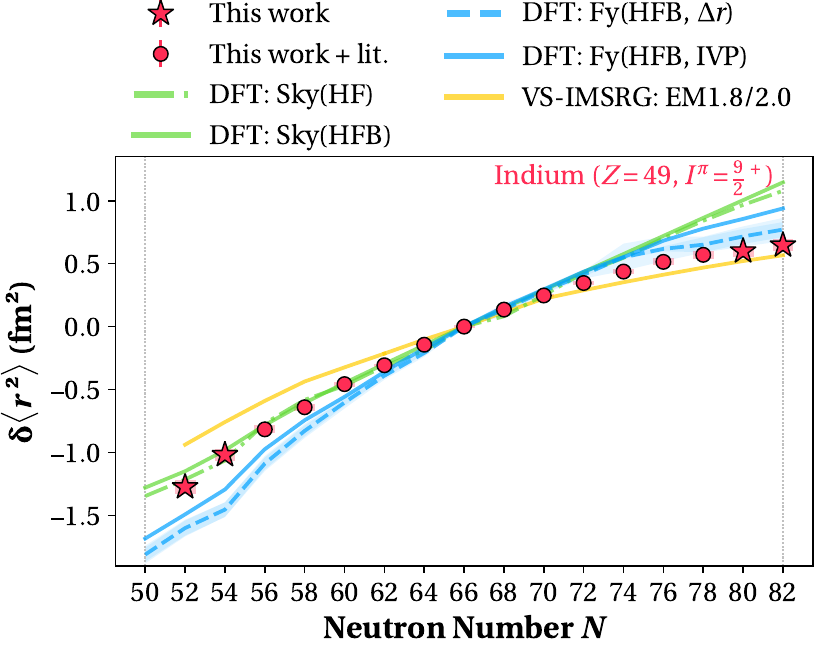}
    \caption{Differential charge radii $\updelta\langle r^2\rangle$ of odd indium isotopes between the two magic neutron numbers $N=50$ and 82. $I^\pi=9/2^+$ ground states with respect to the charge radius of $^{115}$In\cite{dr2ref} are shown as a weighted average of both transitions from this work (red stars) and including the two recalculated transitions from Ref.~\cite{Ebe87} (red circles), if available, (for details, see Methods-\ref{sec:M-F-sys}) with experimental (thin; based on the weighted variance in case of indium; see Methods-\ref{sec:M-F-sys}) and theory-derived (broad error bars) standard deviations and compared to theoretical calculations (colored lines; with (darker) systematic and (lighter) statistical standard deviation bands, if applicable).}
    \label{ExtDatFig:1:dr+theory}
\end{figure}
\renewcommand{\thefigure}{2}
\begin{figure}
    \includegraphics[width=0.5\linewidth]{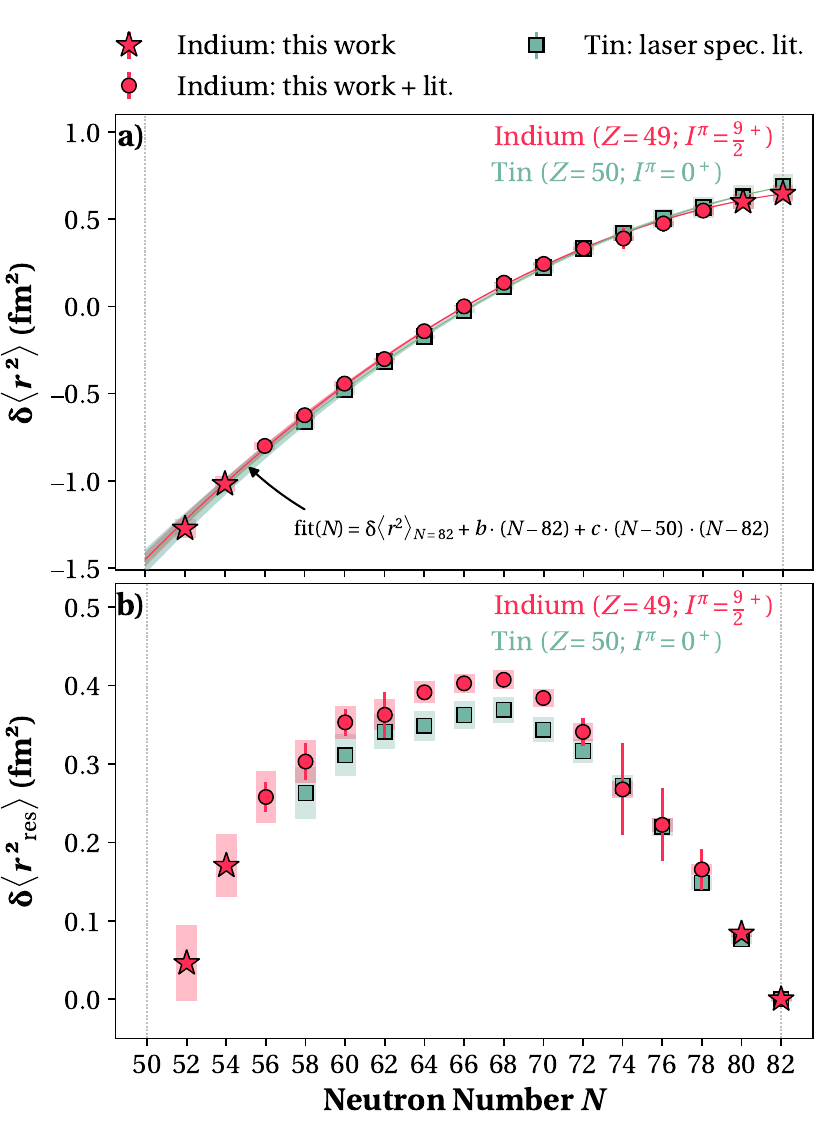}
    \caption{Deduction of relative differential charge radii: a) Differential charge radii $\updelta \langle r^2 \rangle$ for tin (green square markers; Ref.~\cite{Gor19} for $N<58$ and Ref.~\cite{Gor19} for $N>56$) with respect to the charge radius of $^{116}$Sn$(N=66)$~\cite{dr2ref} and indium with respect to the charge radius of $^{115}$In$(N=66)$~\cite{dr2ref} as a weighted average of both transitions from this work (red stars) and including the two recalculated transitions from Ref.~\cite{Ebe87} (red circles), if available, (for details, see Methods-\ref{sec:M-F-sys}). The parabolic fits were fixed to the differential charge radii at $N=82$ and shown with their MC-based one-standard-deviation uncertainty bands. b) Residual differential charge radii $\updelta \langle r^2_{\rm{res}} \rangle$ after subtraction of the linear trend between the fit results in a) at $N=50$ and $N=82$ shown with their experimental (thin; based on the weighted variance in case of indium; see Methods-\ref{sec:M-F-sys}) and theory-derived (broad error bars) standard deviations.}
    \label{ExtDatFig:2:dr2-fit}
\end{figure}
\renewcommand{\thefigure}{3}
\begin{figure}
    \includegraphics[width=0.5\linewidth]{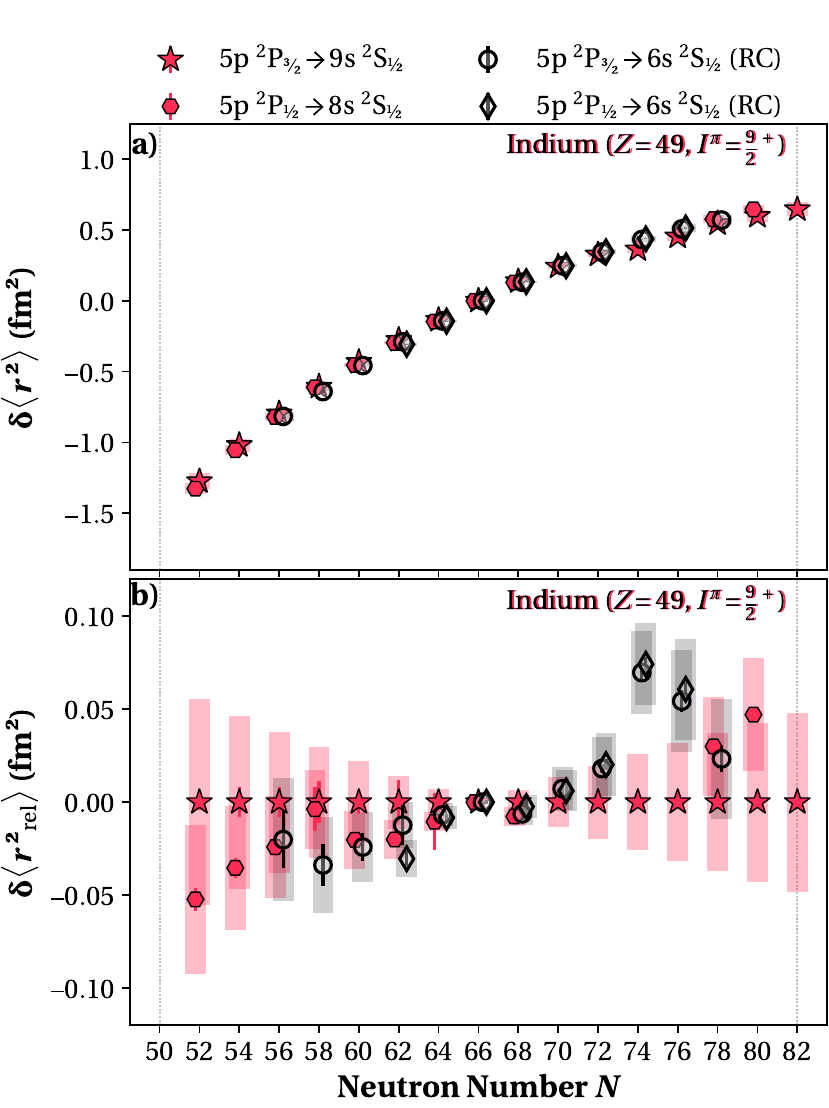}
    \caption{Systematic shifts in differential charge radii of different transitions: a) Differential charge radii $\updelta \langle r^2 \rangle$ for different transitions of $I^{\pi}=9/2^+$ indium ground states (red stars and hexagons for this work; black circles and diamonds for recalculated isotope shifts from Ref.~\cite{Ebe87} based on our atomic theory results). b) Relative differential charge radii $\updelta \langle r^2_{\rm{rel}} \rangle$ for the same transitions after subtraction of this work's 5p$^2$P$_{\nicefrac{3}{2}}$→9s$^2$S$_{\nicefrac{1}{2}}$ transition for improved visibility of the systematic shift. Please note that the markers are slightly shifted toward each other for improved visibility. Experimental uncertainties (thin; based on the weighted variance in case of indium; see Methods-\ref{sec:M-F-sys}) and theory-derived (broad error bars) standard deviations are shown.}
    \label{ExtDatFig:3:shift}
\end{figure}
\renewcommand{\thefigure}{4}
\begin{figure}
    \includegraphics[width=0.5\linewidth]{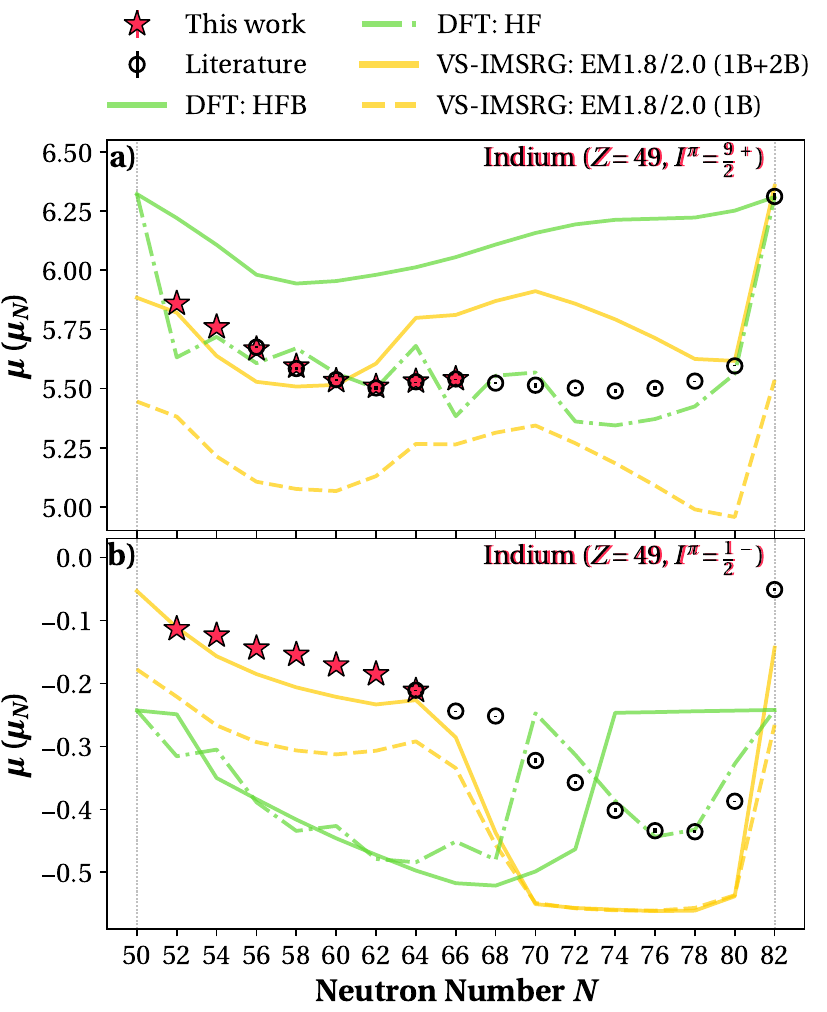}
    \caption{Nuclear magnetic moments $\mu$ of odd indium isotopes between the two magic neutron numbers $N=50$ and 82. $I^\pi=9/2^+$ ground states in panel a) and $I^\pi=1/2^-$ isomers in panel b) from this work (red stars) are compared to the weighted average, if available, of the literature values (black circles) from Ref.~\cite{Ebe87} ($^{105-127}$In for $I^\pi=9/2^+$ and $^{113-125}$In for $I^\pi=1/2^-$) and Ref.\cite{Ver22} ($^{113-131}$In) with their respective standard deviations and theoretical calculations (colored lines). For details, see text.}
    \label{ExtDatFig:4:mu}
\end{figure}
\renewcommand{\thefigure}{5}
\begin{figure}
    \includegraphics[width=0.5\linewidth]{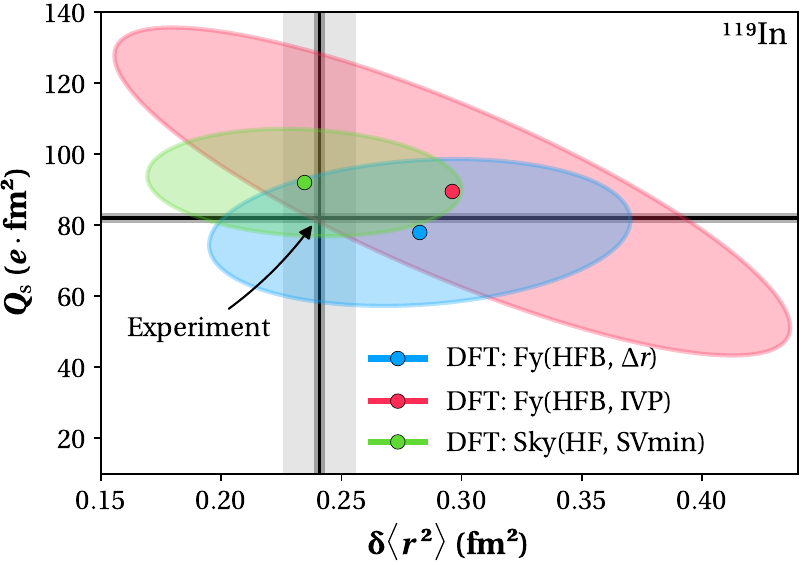}
    \caption{Correlation studies in density functional theory: Typical correlation between the spectroscopic quadrupole moments $Q_{\rm{s}}$ and differential charge radii $\updelta\langle r^2\rangle$ for different DFT parametrizations shown as ellipses and compared to experimental results (vertical and horizontal black lines with grey bars as their uncertainties). The systematic errors from the approximate angular momentum projection have been incorporated. For details, see text.}
    \label{ExtDatFig:5:DFT-corr}
\end{figure}
\renewcommand{\thefigure}{6}
\begin{figure}
    \includegraphics[width=0.5\linewidth]{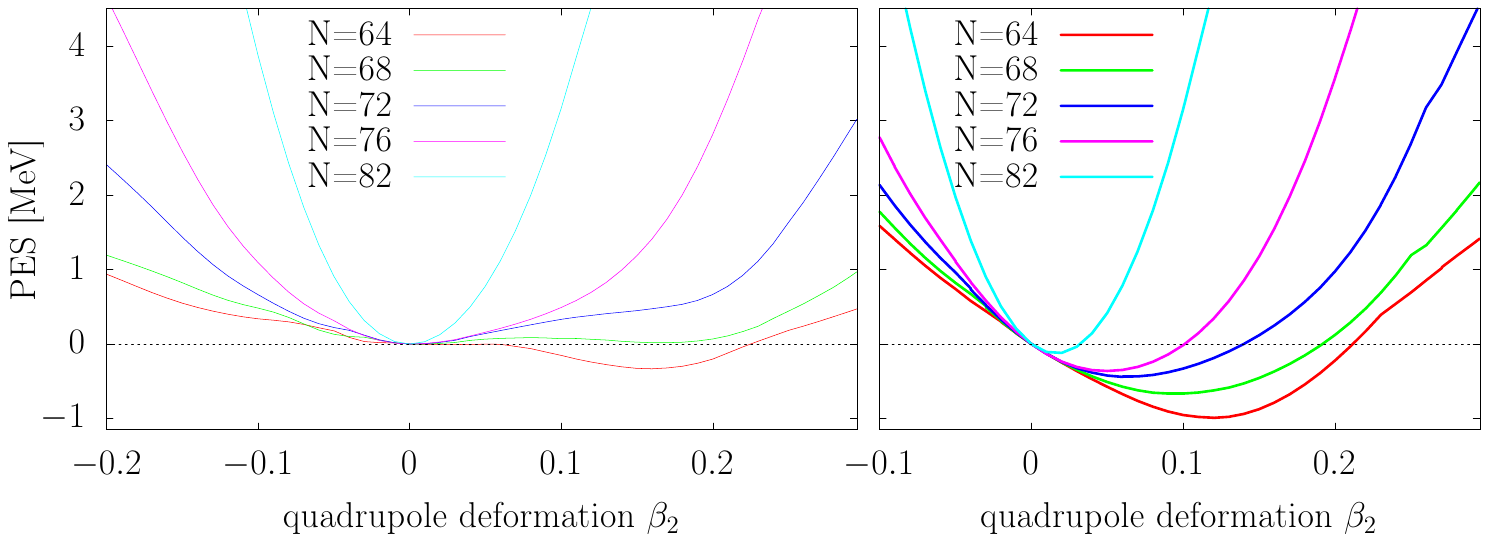}
    \caption{Potential energy surfaces (PES) for selected In and Cd isotopes. The even-even Cd isotopes (left panel) with $N<76$ have extremely soft potential energy surfaces (PES) without well-developed deformed minima. In In isotopes (right panel), in contrast, the PES exhibit well-defined axial-prolate minima for all isotopes. For details, see text.}
    \label{ExtDatFig:6:triax}
\end{figure}

\newpage\hbox{}\thispagestyle{empty}\newpage
\newpage\hbox{}\thispagestyle{empty}\newpage
\newpage\hbox{}\thispagestyle{empty}\newpage

\section*{Methods References}

\end{document}